\documentclass[twocolumn,showpacs,preprintnumbers,amsmath,amssymb]{revtex4}
\usepackage{graphicx}
\usepackage{dcolumn}
\usepackage{bm}
\usepackage{tensor}

\begin{document} 
\title{Coherent chirped pulse laser network \\ in Mickelson 
phase conjugating configuration.}
\author{A.Yu.Okulov}
\email{alexey.okulov@gmail.com}
\homepage{https://sites.google.com/site/okulovalexey}
\affiliation{Russian Academy of Sciences, 
119991, Moscow, Russia}

\date{\ March 1, 2014}
 
\begin{abstract}
{The mechanisms of nonlinear phase-locking of 
a large fiber amplifier array are 
analyzed. It is shown that  
Michelson phase conjugating configuration with double passage through 
array of fiber amplifiers have the definite advantages 
compared to one-way fiber array coupled in a 
Mach-Zehnder configuration. Regardless to amount of 
synchronized fiber amplifiers Michelson phase-conjugating 
interferometer is expected to do a perfect compensation 
of the phase-piston errors and collimation of backwardly 
amplified fiber beams on entrance/output beamsplitter. 
In both configurations 
the nonlinear transformation 
of the stretched pulse envelope due to gain saturation is capable to 
randomize the position of chirp inside envelope 
thus it may 
reduce the visibility of interference pattern 
at output beamsplitter. A certain advantages are inherent to the 
$sech$-form temporal 
envelope because of exponential precursor and self-similar 
propagation in gain medium. The Gaussian envelope is 
significantly compressed in 
a deep gain saturation regime and frequency chirp 
position inside pulse envelope is more deformed.}
\end{abstract}


\maketitle

\section{Introduction}

The chirped pulse laser amplifiers (CPA) aimed to 
generation of a strong optical fields 
and optical 
acceleration of electron and ion beams have attracted 
a special attention from the point view of compactness and 
versatility \cite {Mourou:2013,Tajima:2002}. The possible applications 
range from medical tissue treatment to spallation and further to 
strong field fundamental physics \cite {Bulanov:2006}. 
Recent highly promising trend is a coherent summation of 
the stretched subnanosecond 
laser pulses produced by a thousands of fiber 
amplifiers each operating at millijoule level.
This chirped spatially smooth pulse 
of a tens of Joules may be compressed into femtosecond pulse 
by a diffraction gratings 
\cite {Strickland_Mourou:1985}. The advantage of this 
approach is in massive parallelism which is possible 
due to commercially available thus carefully 
tested $30-50{\:}fs$ fiber 
master oscillators, millijoule level short 
length ($L_f \sim 1 -10{\:} m$) fiber amplifiers and 
reliable fiber beam splitters developed in recent decades 
due to telecommunications needs.

The most impressive project \cite {Mourou:2013} 
presumes the usage of $30fs$ fiber 
master oscillator emitting $10^{-6} J$ pulses with a 
tens of kilohertz 
repetition rate and fiber stretcher in order to produce $\sim 900 ps$ 
frequency chirped pulse. The elongation of pulse 
in time domain is 
necessary to eliminate self-focusing 
and other nonlinear effects during amplification process. 
The Bespalov-Talanov filamentation 
instability \cite {BT_Jetp_lett:1966} 
is suppressed because of smallness 
of the fiber's core diameter ($D \sim 10-120 \mu km$) compared 
to the most dangerous 
filament transverse size in rare-earth doped laser amplifiers 
$ \ell_{\bot} =(2 \pi k \sqrt{2 n_2 I(z,t)})^{-1}
\sim 200-400 \mu km$ 
\cite {Okulov:1988_QE}, where $kn_2 I(z,t)$ is Kerr nonlinearity of 
refractive index induced by optical intensity $I(z,t)$.
Next stage 
involves gradual pulse amplification in a 
multiplexing 
set of standard fiber amplifiers. Each amplifier supports 
single spatial 
mode and $preserves$ frequency chirp, required to compression 
inside the output pair of diffraction gratings. 
The output compressor 
also became a standard component since 
1985 \cite {Strickland_Mourou:1985}. The current 
technology bottleneck  
is a method of the coherent summation 
\cite {Basov:1965} of 
the $N_{f} \sim 10^{(3-4)}$ single spatial 
mode laser beams into a single transversely smooth beam with 
preservation of the frequency chirp to ensure grating compression. 
 
For this purpose  a variety of the linear and 
nonlinear beam coupling techniques 
is being studied \cite {Brignon:2013, Brignon:2011,Tunnermann:2010,
Galvanauskas:2012, Sergeev:2012}. 
The usage of 
$50/50$ beam splitters and Glan-Thompson like polarization cubes 
inside Mach-Zehnder interferometers 
\cite {Tunnermann:2010} (fig.1a) requires the adjustment of the 
optical path difference with an accuracy of $\lambda /(20-50) $ 
and perfect overlapping of elementary 
Gaussian beams. 
Such an adjustment is a quite routine procedure for a low frequency 
thermal and tension noises \cite {Mourou:2013} because 
their noise maximum 
locates below $10 Hz$. Nevertheless despite the 
commercial availability of 
electro-optically controlled phase-shifters and liquid crystal 
light valve wavefront controllers the 
operation of a thousands beamsplitting 
units might overcomplicate a system and 
increase the system cost and operation expenses. The similar 
difficulty is a necessity of providing a high degree of the
spatial overlapping of $10^{(3-4)}$ elementary Gaussian beams 
amplified inside fiber array in a sequence of 
$10^{(3-4)}$ demultiplexing (beam combining) beamsplitters. 
The more serious complications are due to nonlinear 
self-phase modulation (SPM or B-integral 
discrepancy) \cite {Daniault:2012} 
and temporal envelope transformation 
because of gain saturation of laser amplifiers \cite {Basov:1966}. 
The nanosecond laser pulses with exponential 
envelope $sech(t-z n_0/ c)$ move 
in a self-similar style with $superluminal$ envelope speed while 
Gaussian pulses demonstrate self-steepening of precursor and 
$subluminal$ speed with a stronger deformation of envelope. 
We will show that this 
effect is essential for both 
$preservation{\:}{\:} of {\:}{\:}the{\:}{\:}
frequency {\:}{\:}chirp$ 
magnitude and its location within stretched 
pulse envelope $f(t-z n_0/ c)$.

\begin{figure}
\center{ \includegraphics[width=7 cm]{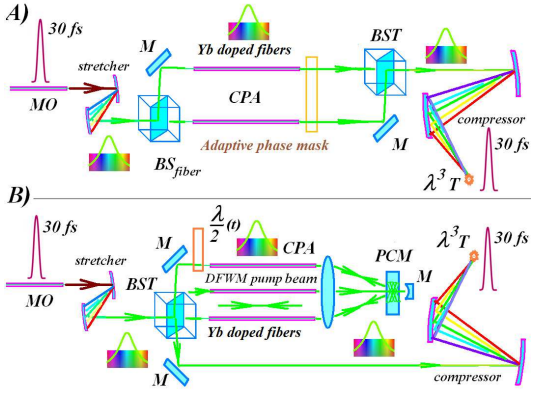}}
\caption{(Color online) A) Chirped pulse  
amplification and compression 
in Mach-Zehnder configuration. ${\bf MO}$ is master ocillator, 
${\bf CPA}$ is chirped pulse amplifying array, 
${\bf BS_{\it{fiber}}}$ \cite {Mourou:2013} 
is entrance beamsplitter tree, which may be 
both free space and fiber array, ${\bf M}$ are ordinary 
retro-mirrors, ${\bf BST}$ is free-space output $binary{\:}{\:} BS$ 
tree \cite {Brignon:2011,Galvanauskas:2012},
 $\lambda^3 \bf T$ is target 
of volume $\lambda^3$. B) Chirped pulse amplification 
and compression in Michelson configuration. 
${\bf PCM}$ is degenerate four - wave mixing (DFWM) 
phase - conjugating mirror, ${\bf {\lambda (t)/ 2}}$ is 
Pockels time-dependent chirp preserving decoupler, ${\bf BST}$ 
is free-space entrance $binary{\:}{\:} {\bf BS}$ tree 
dividing ${\bf MO}$ beam into 
$N_f=2^{N_{ex}} $ amplified beams. Backward phase-conjugated 
amplified emission is combined into single beam 
by this Michelson ${\bf BST}$.}
\label{fig.1} 
\end{figure}
 
In this work we theoretically analyze the coherent beam 
summation technique inside binary tree of beamsplitters 
( $binary{\:}{\:} BST$) without $\lambda/(20-50)$ accuracy adjustment 
\cite {Brignon:2013, Brignon:2011,Tunnermann:2010,
Galvanauskas:2012}. This is possible 
with the aid of phase-conjugating hence self-adjusting 
interferometry 
\cite{Basov:1980, Okulov_2:1980, Pepper:1980,
Boyd:1998, Carroll:1992}. 
The diffraction will be taken 
into account within framework of the 
split-step factorizable model 
\cite {Siegman:1975,Okulov:1988}. Propagation 
inside single Gaussian mode fibers and paraxial 
free space intervals is given by conventional exact solutions
\cite {Okulov:1993,Agrawal:2007}. 
The gain inside fibers 
and nonlinear transformation elements will 
be considered in a 
ray approximation \cite {Okulov:1988}. 
The most attention will be paid to 
phase-locking of fiber array via reflection 
from phase-conjugating mirror (PCM) 
\cite {Basov:1980, Basov_Lett:1980} (fig.4,5). 
In this approach the 
incident signal $E_f (z,t,{\vec r})$ records the 
information about 
gain medium in dynamic hologram written inside 
PCM \cite {Soskin:1990}. 
The reflected phase-conjugated 
signal $E_b (z,t,{\vec r})$ propagates as 
"backward in time" replica with appropriate recovery 
into the smooth single spatial mode beam 
inside a sequence of the Michelson beam splitters (fig.1b). 
The thin slice PCM with instantaneous response inside a 
third order $\chi^{(3)}$ 
Kerr dielectric medium \cite{Yariv:1978,
Zeldovich:1985,Boyd:2007}  
will be considered as 
a solution compatible with preservation of the temporal 
profile of the chirped pulse. 
The basic difference between Mach-Zehnder and 
Michelson configurations is that $binary{\:}{\:} BST$ is 
placed in opposite parts of amplifier array (fig.1a,b), 
although in both 
cases it is constructive interference inside 
a sequence of beam splitters 
that ensures the coherent lossless output. The Mickelson 
amplitude division $binary{\:}{\:} BST$ 
gives a substantial reduction 
of losses compared to wavefront 
division phase-conjugator 
\cite {Rockwell:1986}. 
 
The paper 
is organized as follows. In section II the model is formulated 
for temporal envelope having frequency chirp inside 
array of fibers each having variable 
gain $G_{mn}$ and different length $L_{mn}$ 
resulting in phase-piston errors 
$\Delta \phi _{mn}$. The section III describes the 
procedure of phase-conjugation with preservation of 
frequency chirp and $\Delta \phi _{mn}$ 
compensation. Section IV is devoted to backward 
nonlinear amplification of the phase conjugated replica in 
fiber amplifiers array. The conditions of the chirp preservation 
are formulated qualitatively for Gaussian 
and $sech$ temporal envelopes (fig.6,7). The asymptotic 
behavior 
of output interference pattern 
for $large {\:} N_f $ is studied 
in section V.  
In concluding section VI the results are formulated.
 
\section{Model formulation}
The features of a single passage Mach-Zehnder 
fiber amplifier network (fig.1a) and double-passage 
Mickelson phase-conjugating amplifier 
network (fig.1b) will be analyzed for electric fields of pulses 
$\mathcal E_{{_{Gs}}}$ and 
$\mathcal E_{{_{Gs}}}$ which 
initially ($z=0$) have transform limited 
temporal envelopes of Gaussian and hyperbolic secant form :  
\begin{eqnarray}
\label{envelope_gauss_secant}
{\mathcal E}_{_{Gs,Se}}(z,t,{\vec r}) \cong  {E^o}
\cdot \exp{\:}(-r^2 /2 D^2) \cdot f_{_{Gs,Se}}(\tau) 
&& \nonumber \\
\cdot \exp {\:}[i \theta_{_{Gs,Se}}(\tau)] 
\cdot \exp{\:}(-i \omega t + i k_z z), {\:} {\:} {\:} 
&& \nonumber \\
\theta_{_{Gs,Se}}(\tau) \cong {kn_2} \cdot L_{str} 
\cdot  {|\mathcal E_{_{Gs,Se}}(z,t,r=0)|^2},{\:}{\:}{\:}{\:}
&& \nonumber \\
f_{_{Gs}}(\tau)= \exp {\:} [-\tau^2/2{\tau_{_{Gs}}^2}], 
f_{_{Se}}(\tau)= sech {\:}[\tau/{\tau_{_{Se}}}], 
{\:} {\:} {\:} {\:} {\:} {\:} {\:}
\end{eqnarray}
which undergoes frequency modulation due to Kerr 
nonlinearity $\chi^{(3)}$. Here $E^o$ is electrical field amplitude 
at stretcher output ($z=L_{str}$), $(z,t,{\vec r})$ 
is coordinate system, collocated with propagation axiz $z$, 
$\tau=t-zn_0/c$, 
${\tau_{_{Gs}}},{\tau_{_{Se}}} \sim 30 - 50 fs$ 
are the pulse durations for MO and 
$\sim 900 ps$ for stretched pulse in amplifiers, 
$k_z \sim k = n_0{\:} \omega /c $ is wavenumber, $n_o$ is 
linear refractive index, $v_g=\partial \omega / \partial k=c/n_0$,
$L_{str}$ is the length of the medium (stretcher)  
having $\chi^{(3)}$ susceptibility, $r=|\vec r|$,
$D$ is radius of Gaussian fundamental transverse mode, 
$\theta_{_{Gs,Se}}(\tau)$ is 
proportional to the breakup integral $B$:  
\begin{eqnarray}
\label{breakup_int}
\theta_{_{Gs,Se}}(\tau) \sim {B}(\vec r=0,\tau)=
{\:} {\:} {\:} {\:} {\:} {\:} {\:}
{\:} {\:} {\:} {\:} {\:} {\:} {\:}
&& \nonumber \\
{\frac {2\pi n_2}
{\lambda}}
{\int_{_{_{0}}}^{L_{str}}}|{\mathcal E}_{{_{Gs, Se }}}
(z,\tau,r=0) |^2 d z, {\:}{\:} n_2 = 
\frac {3 \chi^{(3)}}{8 n_0}, {\:}
 {\:} {\:} {\:} {\:} {\:} {\:} {\:}  {\:} {\:}
\end{eqnarray}
calculated as a nonlinear phase accumulated over 
distance $L_{str}$ for a given moment $\tau$ of envelope 
propagating along fiber 
axis ($r=0$),${\:}{\:} n_2 |{{ E}_{{_{Gs, Se }}}}|^2$ is 
Kerr refractive index, $n_2$ is $\sim 10^{-20}m^2/W$ for typical 
glasses, $n_2 \sim 10^{-19}m^2/W$ for crystals alike Nd:YAG and even 
more for resonant media e.g. sodium vapor,  
$n_0$ is linear refractive index of stretcher. This oversimplified 
estimate of self-phase modulation is adequate for short Kerr 
medium where group velocity dispersion 
$\beta={{\partial^2 k}/{\partial \omega^2}}$ is negligible  
in nonlinear Shrodinger equation (NLS) for slowly varying 
envelopes ${E}_{_{Gs,Se}}(z,\tau)$: 
\begin{eqnarray}
\label{NL_Shrodinger}
i{\frac {\partial { E}_{_{Gs,Se}}(z,\tau)}{z}} = 
\frac {\beta}{2} \cdot 
{\frac {\partial^{2} { E}_{{_{Gs,Se}}}
(z,\tau)}{\partial^2 \tau}} 
- \gamma {|{E}_{{_{Gs,Se}}}(z,\tau)|^2} \times 
&& \nonumber \\
{E}_{{_{Gs,Se}}}(z,\tau), {\:}{\:}
A_{eff}= \frac {[{\:}\int \int_{-\infty}^{\infty} 
|{E_{_{Gs,Se}}}|^2 d^2 {\vec r}{\:}{\:}]^2} 
{{\int \int_{-\infty}^{\infty} |{E_{_{Gs,Se}}}|^4 d^2 
{\vec r}}}{\:}{\:},{\:}{\:}{\:}{\:}
\end{eqnarray}
where $\gamma = (n_2 \omega) /(c A_{eff})$, $A_{eff}$ is 
effective area of a fiber. Frequency modulation 
in each given moment $t$ is provided by 
$\delta \omega_{{_{Gs,Se}}}(\tau)= 
{\partial {\theta_{{_{Gs,Se}}}(\tau)}/{\partial \tau}}$. 
The phase factors $\theta_{{_{Gs,Se}}}(\tau)$ 
as a function of a local 
time $\tau = t - z/v_g$ are shown in fig.4a for $sech$ envelope 
(eq.) and for Gaussian envelope (eq.)at fig.4b. In both cases 
the linear frequency chirp appears near the 
parabolic maximum of 
pulse envelope. 
	The split-step approach to NLS means the separate integration 
of nonlinear phase-modulation and dispersion terms. Let us split NLS 
in two slices only. This removes from consideration the 
possible effects 
of temporal solitons formation \cite{Agrawal:2007} 
and their interaction during pulse 
stretching in fiber from femtosecond to nanosecond duration.
Instead the smooth analytical formulas 
relevant to temporally elongated nanosecond pulse are 
used. The remained integration is due to group 
velocity dispersion (GVD) $\beta=\partial^2 k /\partial \omega ^2$: 
\begin{equation}
\label{linear_Shrodinger}
i{\frac {\partial { E}_{_{Gs,Se}}(z,\tau)}{z}} = 
\frac {\beta}{2} \cdot 
{\frac {\partial^{2} { E}_{{_{Gs,Se}}}(z,\tau)}
{\partial^2 \tau}},
\end{equation}
The solution via Fourier transform 
of initial value (Cauchy) problem from $z=0$ to 
any $z>0$ is given by:
\begin{equation}
\label{solution_fourier}
{ E}_{_{Gs,Se}}(z,\tau)= 
\int_{-\infty}^{\infty} 
{ {\frac {{ E}_{_{Gs,Se}}(z=0, \omega) }{2 \pi }} 
\exp [\frac {i}{2} \beta \omega^2 z - i \omega \tau]}{\:} d {\omega},
\end{equation}
where spectra of Gaussian and $sech$ envelopes at output facet 
of master oscillator are given by: 
\begin{eqnarray}
\label{spectra}
{E}_{_{Gs}}(z=0,\omega)= 
\int_{-\infty}^{\infty} 
\exp {\:} [-\tau^2/{2 \tau_{_{Gs}}^2}]
\exp [ i \omega \tau]{\:} d {\tau}= 
&& \nonumber \\
{\frac {{\tau_{_{Gs}}}{\sqrt \pi}}{2}}
\exp[-\omega^2 {\tau_{_{Gs}}^2}/4], {\:}{\:}
&& \nonumber \\
{ E}_{_{Se}}(z=0,\omega)=
\int_{-\infty}^{\infty}
sech {\:} [\tau/{\tau_{_{Se}}}]
\exp[ i \omega \tau]{\:} d {\tau}=
&& \nonumber \\
\frac {\pi {\tau_{_{Se}}}}{2}
sech(\frac{\pi \omega {\tau_{_{Se}}}}{2}).{\:}{\:}{\:}{\:}
\end{eqnarray}

The broadening of pulses due to GVD $\beta$ is given by:
\begin{eqnarray}
\label{spectra_exact}
{ E}_{_{Gs}}(z,\omega)= (2 \pi \tau_{_{Gs}})^{1/2}
\exp {\:} [-\omega^2 \tau_{_{Gs}}^2/4 + i \omega^2 {\beta z}/2]
, {\:}{\:}
&& \nonumber \\
{ E}_{_{Se}}(z,\omega)\sim 
sech {\:} [\pi \omega {(\tau_{_{Se}}-\beta \omega z)/2}].
{\:}{\:}{\:}{\:}
\end{eqnarray}
Both cases demonstrate the self-similar stretching due 
to propagation along $z$. Propagation of chirped pulses 
is somewhat more complicated due to nonlinear phase $\theta (\tau)$:
\begin{eqnarray}
\label{spectra chirped initial}
{ E}_{_{Gs}}(L_{str},\omega)\cong
\int_{-\infty}^{\infty} 
\exp {\:} [-{\frac {\tau^2}{2 \tau_{_{Gs}}^2}}]
\exp [ i \omega \tau+ i\theta_{_{Gs}} (\tau)]{\:} d {\tau}, 
&& \nonumber \\
{ E}_{_{Se}}(L_{str}, \omega)\cong 
\int_{-\infty}^{\infty}
sech {\:} [\tau/{\tau_{_{Se}}}]
\exp[ i \omega \tau + i\theta_{_{Se}} (\tau)]{\:} d {\tau}.
{\:}{\:}{\:}{\:}{\:}{\:}
\end{eqnarray}
The explicit formulas for the chirped spectra of 
the ${ E}_{_{Gs,Se}}(z,\omega)$ for $z>L_{str}$ 
for the short (quasiclassical) distance $z$ 
are obtained 
via stationary phase method \cite {Okulov:1988,Rytov:1987} 
because $\theta_{_{Gs,Se}}(\tau)$ has a slowly 
varying parabolic extremum near $\tau=0$. 
The spectrum for Gaussian pulse is modified as:  
\begin{eqnarray}
\label{spectra gauss chirped stretched}
{E}_{_{Gs}}(z,\omega)\cong 
{\tau_{_{Gs}}}\sqrt {\frac {{2 \pi}}
{2 {kn_2} \cdot 
L_{str} {\:} \partial^{{\:}2}{E}_{{_{Ga}}}(0,\tau =0)/\partial {\tau}^2}}
{\:} \times 
&& \nonumber \\
{\:}{\:}{\:}{\:}{\:}{\:}
{\:}{\:}{\:}{\:}{\:}{\:}{\:}
\exp[-\omega^2 {\tau_{_{Gs}}^2}/4]{\:} \times
&& \nonumber \\
\exp [ \frac {i}{2} \beta \omega^2 z 
+ i {kn_2} \cdot L_{str} 
\cdot |{E}_{{_{Ga}}}(0,\tau=0)|^2 +i \pi /4]{\:} , {\:}{\:}
\end{eqnarray}
whereas for $sech(\tau)$ pulse $z-dependent$ spectrum is:
\begin{eqnarray}
\label{spectra sech_chirped stretched}
{ E}_{_{Se}}(z, \omega)\cong 
{\tau_{_{Gs}}}\sqrt {\frac {{2 \pi}}
{2 {kn_2} \cdot 
L_{str}{\:} \partial^{{\:}2}{E}_{{_{Se}}}(0,\tau=0)/\partial {\tau}^2}}
 {\:}\times 
&& \nonumber \\
sech(\frac{\pi \omega {\tau_{_{Se}}}}{2}) \times 
{\:}{\:}{\:}{\:}{\:}{\:}{\:}{\:}{\:}{\:}{\:}{\:}{\:}{\:}
&& \nonumber \\
\exp[ \frac {i}{2} \beta \omega^2 z 
+ i {kn_2} \cdot L_{str} 
\cdot |{E}_{{_{Se}}}(0,\tau=0)|^2 +i \pi /4]{\:} .{\:}{\:}{\:}{\:}
\end{eqnarray}

The recovered temporal envelopes of nanosecond 
duration ${ E}_{_{Gs,Se}}(z,\tau)$ have a 
linear frequency modulation (chirp) near pulse maximum in such 
a way that precursor is red-shifted with respect to the 
pulse tail \cite {Strickland_Mourou:1985}. The same frequency 
chirp is produced by a pairs of diffraction gratings, 
prisms with chromatic aberrations  and grisms (prisms with grooved 
gratings). 

Let us extend now the model for phase locking of 
multiple beams amplified in fibers or bulk crystals.  
Using temporal envelops ${E}_{_{Gs,Se}}(z,\tau)$
the high frequency field $\mathcal E_{mn} $ in 
each fiber channel ${(m,n)}$ is: 
\begin{eqnarray}
\label{factorized}
{\mathcal E}_{mn}(z,t,{\vec r}) \cong  {E^o}_{mn} 
\cdot \exp{\:}[-(\vec r-\vec r_{mn})^2 /2 D^2] \cdot 
&& \nonumber \\
{E}_{_{Gs,Se}}(z,\tau)
\cdot \exp{\:}(-i \omega t + i k_z z + 
i \Delta \phi_{mn}), {\:} {\:} {\:} 
\end{eqnarray}

Consider the spatially periodic lattice with 
period $p \sim 2D$ composed of the polarization 
preserving fiber laser amplifiers 
whose output facets are located at 
points $\vec r_{mn}$ , where $\vec r$ is 
transverse coordinate, 
$z$ is coordinate along pulse propagation direction, $t$ - is time. 
The linearly polarized output 
field of this array $\mathcal E_f(z,t,{\vec r})$ is a superposition 
of the elementary partial waves each having transversely Gaussian 
profile with radius $D$ \cite{Okulov:2008a, Okulov:1993}: 
\begin{eqnarray}
\label{fiber array}
{\mathcal E}_f(z,t,{\vec r}) \cong 
\exp[-i \omega t + i k_z z ]
\cdot{ E_{_{Gs,Se}}(z,\tau)}
&& \nonumber \\
\sum_{m,n}  E^o_{mn} 
\cdot \exp[-(\vec r - \vec r_{mn})^2 /2 D^2+i \Delta \phi_{mn} ],
\end{eqnarray}
where $E^o_{mn}$ is electric field amplitude, 
$E_{_{Gs,Se}}(z,\tau)  = f_{_{Gs,Se}} (t-z/v_g  ) \cdot 
exp {\:}  [i \theta_{_{Gs,Se}} (t-z/v_g )]$ is 
temporal envelope function,
$\Delta \phi_{mn}$ is a random phase $piston$ shift 
labelled by indices ${(m,n)}$, induced 
by fiber's length variation $L_{mn}$, heating 
or stress inside  a given fiber ${(m,n)}$ amplifier 
\cite {Okulov:1991}. 

\section{Phase-conjugation of laser array emission with chirped pulse 
temporal envelope}
 
The accurate control of the phase piston errors $\Delta \phi_{mn}$ 
with accuracy about $\lambda / (20-50)$ is well studied for the small 
number ($N_f \cong 2 - 4$) of phase-locked fiber 
amplifiers and a reasonable energy efficiency 
(more than 95 percents) had been reported 
already \cite {Tunnermann:2010}. In most cases the 
Mach-Zehnder interferometry 
was used because it ensures temporal envelope preservation 
for $\lambda/(20-50)$ path difference between 
synchronized beams. The question is in robustness 
of this technique for a large number $N_f \cong 2^{N_{ex}}=
4096 - 32768$ of a phase locked laser amplifiers. 
Here $ N_{ex} \sim 12-15$ is 
a number of superpositions experienced by 
each elementary beam inside  $binary{\:}{\:} BST$ to achieve 
perfect smooth beam combination. The total 
number of required beamsplitters 
$N_{bs}$ grows $linearly$ when $N_f$ 
is increased.  
Starting from $N_f=2$ when one beamsplitter is 
needed ($N_{bs}=1$) 
it is clear that network which combine $N_f = 2^{N_{ex}}$ 
fiber lasers 
contains $N_{bs}=(N_f/2)/(1-0.5)=N_f-1$ beamsplitters 
(sum of $geometric$ $progression$). 
This network requires not only careful adjustments of the phase lags 
$\Delta \phi_{mn}$ before each beam splitter. The other urgent requirement 
is to ensure perfect  
spatial overlapping of transverse Gaussian profiles of the each beam 
at the each of $N_f-1$ beamsplitters 
of the coherent beam summation network. 

\begin{figure}
\center{ \includegraphics[width=7 cm]{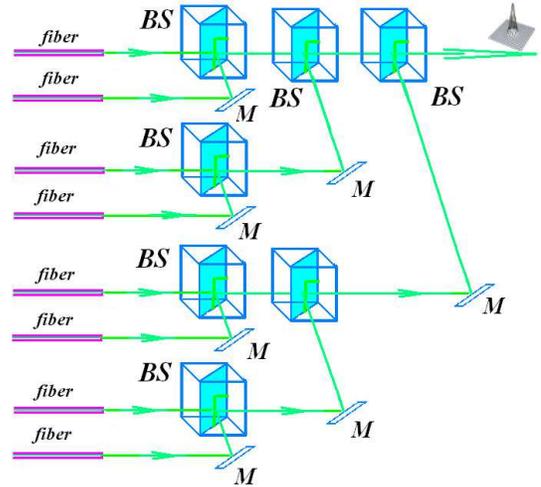}}
\caption { (Color online) Number of beams splitters BS required 
for coherent summation of $N_f$ = 8 fiber amplifiers. 
This binary tree placed at output of Mach-Zehnder interferometer 
(fig.1a ) is a pyramid with ratio of 
elements in adjacent layers 0.5. As it easily seen 
by summation of the clearly visible geometric progression, 
number of BS and mirrors M required is exactly $N_f$ - 1.}
\label{fig.2} 
\end{figure} 

This task 
would require an intense usage of the micropositioners and 
substantial computing resources. The preliminary evaluation 
of computing and micropositioning resources might be performed 
in the following way. Indeed each of the $N_{bs}$  
beamsplitters has $six$ degrees of freedom. In addition position of 
the each elementary Gaussian beam 
directed to combining beamsplitter is controlled by two 
tranverse coordinates , two angles and focal point location. 
Thus the upper 
bound on the total amount 
of dynamical variables to be controlled in Mach-Zehnder 
network (fig.1a) 
is $N_{var}=(N_f-1)*6 + 2*(N_f-1)*(2+2+1)$. 

\begin{figure}
\center{ \includegraphics[width=7 cm]{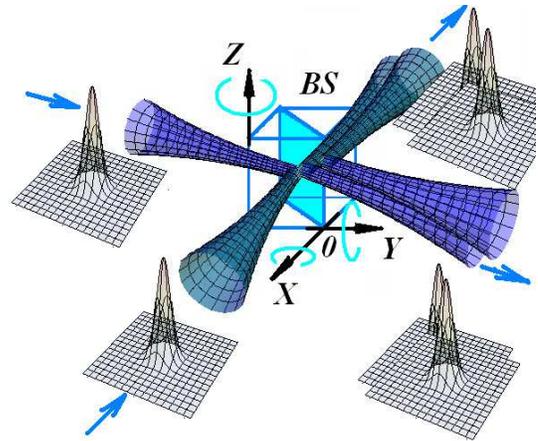}}
\caption{(Color online) Number of degrees of freedom 
controlled for ensured constructive interference in BS.  
Gaussian beams (which have hyperboloid isosurfaces ) 
at entrance ports of BS are controlled by two 
transverse coordinates, two paraxial angles and 
one longitudinal parameters (distance towards Gaussian beam waist  ).  
Shifted Gaussians in output ports form 
interference pattern similar to the two plane 
wave intersection or "equal tilt fringes". }
\label{fig.3} 
\end{figure}
The alternative method of phase locking of 
multiple beams amplified in fibers or bulk crystals 
is Michelson phase-conjugator (fig.1b) which looks 
experimentally attractive compared to 
Mach-Zehnder interferometric schemes (fig.1a) 
\cite {Galvanauskas:2012,Tunnermann:2010, Brignon:2011}. 
In addition to currently studied nonlinear beam coupling 
techniques with one pass propagation 
which use second order parametric 
processes with $\chi^{(2)}$ nonlinearities 
\cite {Sergeev:2012} 
or four-wave 
mixing $\chi^{(3)}$ processes we consider double-pass configuration 
\cite {Basov:1980} which is a well-proven 
tool for compensation of the phase-piston errors 
$\Delta \phi_{mn}$ in amplifying 
channels \cite {Rockwell:1986}. 

For the first sight stimulated Brillouin scattering (SBS) 
looks feasible 
for phase-conjugation of chirped laser pulses of nanosecond duration. 
The conceptual difficulty is in accuracy of reproduction of 
the temporal envelope $f(t-z/v_g  )$ \cite {Okulov:1983}. For the 
long chirped pulse 
$\tau_{_{Ga,Se}}>\tau_{ph} \sim 10^{-9}sec$ the 
phase modulation (linear in time 
frequency chirp) 
will be distorted 
by a random phase jumps separated 
by interval $\sqrt {2 \pi G_{_{sbs}}} {\:} \tau_{ph}$ caused 
by finite lifetime of 
acoustical phonons $\tau_{ph}=1 / \Gamma $ 
\cite {Basov_Lett:1980}. Here 
$G_{_{sbs}} \cong  25-30 $ is SBS increment (gain 
growth rate), $\Gamma=2 \eta {(k_p+k_s)^2 } /3 \rho_0 $ is sound damping 
rate (spontaneous Brillouin 
scattering linewidth) \cite {Zeldovich:1985,Okulov:1983}, $\eta $ is 
viscosity of Brillouin medium.
On the other hand for the short nanosecond 
laser pulses having $\tau_{_{Gs,Se}} \sim 10^{-9}sec$ SBS 
reflected wave $E_b (z,t ,\vec r)$ is not 
able to follow the frequency modulation of the pump wave 
$E_f (z,t ,\vec r)$  
because of inertia of acoustic wave $Q (z,t ,\vec r)$.

Indeed SBS equations of motion for the scalar slowly varying envelope 
optical fields, i.e. ${{ {E}}}_{f}$ moving in the 
positive $z$-direction and ${ {E}}_{b}$ moving oppositely are:
\begin{equation}
\label{pumpwave}
\ {\frac {\partial {{{{{E}}}_f}(z,t,\vec r )}} {\partial z} }+
{\frac {n_0} {c} }{\frac {\partial {{{{{E}}}_f}}} 
{\partial t}}+
{\frac {i}{2 k_p}} {\nabla}_{\bot}^{{\:}2} {{{{E}}}_f} =
{\frac {i \gamma_{_{SBS}} \omega_p {\:}} 
{4{\:} c{\:}{n_0} \rho_0 } } { Q} {\:}{{ {E}}}_b 
\end{equation}
\begin{equation}
\label{stockeswave}
\ {\frac {\partial {{{{ {E}}}_b}(z,t,\vec r )}} {\partial z} }-
{\frac {n_0} {c} }{\frac {\partial 
{{{{ {E}}}_b}}} {\partial t} }-
{\frac {i}{2 k_s}} {\nabla}_{\bot}^{{\:}2} {{{ {E}}}_b} = -
{\frac {i \gamma_{_{SBS}} \omega_s {\:}} {4{\:} c{\:}{n_0} \rho_0 } } 
{{ {E}}}_f {{Q}}^{\ast},
\end{equation}
with dimensionless slowly varying acoustical 
perturbation complex 
amplitude ${ Q}$ 
\cite{Zeldovich:1985}: 
\begin{equation}
\label{acouswave1}
\ v_{ac} {\frac {\partial {{Q}(z,t,\vec r )}} {\partial z} }+
{\frac {\partial { Q}}  {{\:}\partial t} }
+{\frac {{\Gamma} {{Q}}} {2 } }=
 {{{ {E}}}_f}  
{{{ {E}}}_b }^{\ast}{\frac {{i \gamma_{_{SBS}} 
(k_p+k_s)^2 }}{16 \pi{\:}{\:}\omega_{ac}}},
\end{equation}
where 
$\gamma_{_{SBS}}=\rho {\:}{\:}(\partial \epsilon / \partial \rho)_S$ 
is electrostrictive coupling constant \cite {Okulov:2008J}, 
$\rho$ is density 
of SBS medium,
$n_0$ is refractive index, $v_{ac}$ is speed of sound. 

As is shown experimentally and computationally in many 
works the phase of reflected 
Stockes pulse in the limit $\tau_{_{Ga,Se}} >>\sqrt {2 \pi G_{_{sbs}}} \tau_{ph}$ 
experiences a random phase 
jumps $\Delta \Phi_{Stockes}$ uniformly distributed 
in the interval $(-\pi,\pi)$ \cite {Basov_Lett:1980}. 
For the broadband SBS pump radiation when 
characteristic correlation time is much shorter than lifetime 
of acoustic phonons $\tau_{ph}$ the Stockes wave envelope is 
modulated by random phase jumps $\Delta \Phi_{pump}$ 
of pump wave $E_f(z,t)$  \cite {Okulov:1983}. 
In such a case linear 
frequency chirp is not reproduced by SBS mirror. The short 
pulse  
limit $\tau_{_{Ga,Se}} << \sqrt {2 \pi G_{_{sbs}}} \tau_{ph}$ means 
the fixed carrier 
frequency of acoustical phonons. Consequently the carrier 
frequency 
difference $\omega_{ac}=\omega_{f}-\omega_b$ between pump 
and Stockes photons do $not$ feel the frequency chirp of incident 
wave $E_f(z,t)$. 
Hence phase-conjugated replica $E_b(z,t)$ having 
carrier frequency $\omega_b$ can not 
be compressed after reflection from SBS PCM.  

For this reason the usage of phase-conjugating mirror with 
$instantaneous$ response e.g. mirror based 
on $\chi^{(3)}$ Kerr-like instantaneous 
nonlinearity \cite{Yariv:1978,Zeldovich:1985} looks attractive.  
The wavefront reversal via degenerate four-wave mixing 
(DFWM) \cite{Soskin:1990} is described 
by nonlinear Shrodinger equation for the  
wide area $\chi^{(3)}$ slice:

\begin{equation}
\label{NL_Shrodinger_DFWM}
{\frac {\partial {E^{PC}_{_{Gs,Se}}(z,t,\vec r)}}{z}} = - 
\frac {i}{2k} \cdot 
 {{\nabla}_{\bot}^{{\:}2}  E^{PC}_{{_{Gs,Se}}}}
- {\frac {i \gamma_{_{PC}}}{2}} {|E^{PC}_{{_{Gs,Se}}}|^2} \cdot
E^{PC}_{{_{Gs,Se}}} {\:}{\:},
\end{equation}

where $E^{PC}_{_{Gs,Se}}$ is a superposition of the 
four optical fields having Gaussian or $sech$ temporal envelope: 

\begin{equation}
\label{DFWM_superposition}
E^{PC}_{_{Gs,Se}}(z,t,\vec r) = E_1+E_2+E_f+E_b
 {\:}{\:},
\end{equation}
where $E_1$  and $E_2=E_1^{*}$ are phase conjugated pump 
beams with $chirped{\:} spectra$ ,
$E_f$ is fiber array output beam, corrugated by phase-piston 
errors $\Delta \phi_{mn}$ (Eq.12), $E_b \sim E_f^*$ 
is phase-conjugated replica 
generated within DFWM PC mirror.  
\begin{figure}
\center{ \includegraphics[width=7 cm]{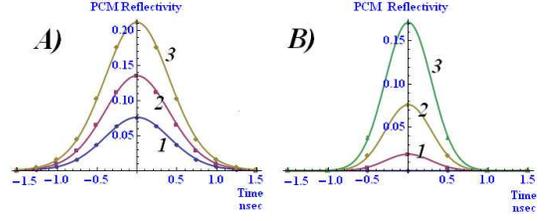}}
\caption{(Color online)Phase conjugation of $sech(t)$ 
and Gaussian chirped pulses in 
degenerate four-wave $\chi^3$ mirror. Reflectivity 
$R_{_{PCM}}(t)\sim  L^2_{_{PCM}} \gamma^2_{_{PC}} 
I^2(z=0,t)\sim \theta^2_{{_{Gs,Se}}}$
is proportional to the square of 
intensity (1-3, gradually increased, arb.units).
 A) $sech^4(t/{\tau_{_{Se}}})$, 
B) is $exp[-4\tau^2/2{\tau^2_{_{Ga}}}]$.}
\label{fig.4} 
\end{figure}
Reflectivity of PC mirror $(R_{_{PCM}})$ and PC fidelity 
$(K_{_{PCM}})$ are  given by conventional 
equation for the four-mixing phase-conjugation 
inside thin Kerr slice
\cite{Yariv:1978,Zeldovich:1985} (fig.2): 
\begin{eqnarray}
\label{Thin_Kerr_DFWM}
{\frac {\partial {E_b}_{_{Gs,Se}}(z,t,\vec r)}{z}} = 
{\frac {i \gamma_{_{PC}}}{2}} E_1  E_2  f^2(\tau)
{{E_f}^{*}_{_{Gs,Se}}(z,t,\vec r)},
&& \nonumber \\
R_{_{PCM}}(t)=\frac{|E_b(z=0,t)|^2}{|E_f(z=0,t)|^2}= f^4(\tau) \cdot 
L^2_{_{PCM}} \gamma^2_{_{PC}}  E_1 \cdot E_2,
&& \nonumber \\
K_{_{PCM}}= \frac {{\:}|\int\int_{-\infty}^{\infty} 
{E_{f}} {E^{*}_{b}} d^2 {\vec r}|^2{\:}{\:}} 
{{\int \int_{-\infty}^{\infty} |{E_{f}}|^2 d^2 {\vec r}}{\:} \cdot
{\int \int_{-\infty}^{\infty} |{E_{b}}|^2 d^2 {\vec r}}} 
{\:}\le 1,{\:}
{\:}{\:}{\:}{\:}{\:}{\:}{\:} 
\end{eqnarray}
where Kerr nonlinearity of DWFM PCM 
is $\gamma_{_{PC}}=k {\:}  3 \chi^{(3)} /8 \pi $, 
$L_{_{PCM}}$ is thickness of the Kerr slice. 
Temporal dependence of 
PCM reflectivity (fig.4) preserves the envelope form near 
pulse maximum and asymptotic dependencies at precursor and 
pulse tail. The undepleted pump reflection 
regime \cite{Yariv:1978, 
Zeldovich:1985,Boyd:2007} with $R_{_{PCM}}\cong 0.2$ 
is considered here 
to avoid development of Bespalov-Talanov 
instability \cite{BT_Jetp_lett:1966, Okulov:1988_QE, Zeldovich:1985}.
\begin{figure}
\center{ \includegraphics[width=7 cm]{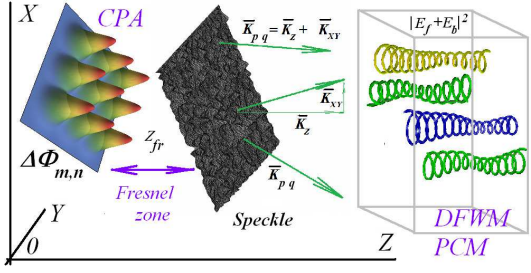}}
\caption{(Color online)Principle of phase locking of the 
CPA fiber array 
with phase-piston errors 
$\Delta \phi_{mn}$ having parameters of 64 fiber 
array \cite {Brignon:2011} with 
$p \sim D \sim 100 \mu m$. Inside Fresnel 
zone ${z_{_{fr}}}\sim D^2/{\lambda}$ the interference of 
overlapping Gaussian beams produces speckle pattern 
\cite {Rytov:1987} 
which is randomly spaced set of field zeros. 
This random field composed of vortex-antivortex pairs
\cite {Okulov:2008J} passes into degenerate four wave mixing 
PC mirror (DFWM PCM).
Each vortex produces helical interference pattern with it's 
phase-conjugated 
replica \cite {Okulov:2009}. Phase conjugated 
field ${E_b}_{Ga,Se}{(z,\tau,\vec r)}$ propagates backward 
along curvilinear helical waveguides 
with $\lambda/2$ modulation and recovers phase-piston 
errors $\Delta \phi_{mn}$. This ensures optimal collection 
of reflected and backwardly amplified field at entrance/output 
beamsplitter $\bf BST$. Each speckle inhomogeneity is considered 
here as a tilted plate emitting plane wave with 
wavevector $\vec K_{pq}$. 
$\vec K_{X,Y}$ is deflection 
of $\vec K_{pq}$ from normal $\vec K_Z$.}
\label{fig.5}
\end{figure}
Experimentally $E_1,E_2$ are 
smooth counter propagating Gaussian beams with phase matched 
wavefronts.
The spatial structure of factorized 
incident $E_f$ and phase-conjugated beams 
$E_b$ is given by a following procedure. 
Let us decompose $E_f$ in Fourier plane-wave series 
with randomly tilted wavevectors $\vec K_{pq}$. 
For the sake 
of computational convenience consider the superposition of 
waves emitted by output 
fiber facets as plane waves. The period of fiber 
array $p \sim 100 \mu m $
is taken comparable to fiber mode at output microlens 
$2 D \sim 100 \mu m$ \cite {Brignon:2011}. 
The resulting interference pattern 
in Fresnel zone, i.e. at the distance 
${z_{_{fr}}} \sim D^2/{\lambda}$ from 
fiber array output plane 
is identical to field which passed through randomly corrugated 
phase-plate (fig.5). 
The 2D Fourier sum of plane waves with "global" wavevectors 
$\vec K= \vec K_z + \vec K_{pq}$ may be reduced 
to the 1D Fourier sum where each 
plane wave with "local" wavevector 
$\vec K= \vec K_z + \vec K_{M}$ 
is emitted by a randomly tilted smooth area 
located at equivalent phase plate in near 
field \cite {Okulov:2009}:
\begin{eqnarray}
\label{speckle_forward}
{\mathcal E}_f(z,t,{\vec r}) \cong 
\exp[-i \omega t + i k_z z +i \theta_{_{Gs,Se}}(\tau)]
&& \nonumber \\
\cdot{f_{_{Gs,Se}}(\tau)}
\sum_{M}  a_{M} 
\cdot \exp[i \vec{K}_M \vec r ],
\end{eqnarray}
where $\vec{K}_M$ is randomly tilted vector of partial 
speckle plane wave, $a_{M}$ is Fourier amplitude, 
$f_{_{Gs,Se}}(\tau)$ is 
temporal envelope. The phase-conjugated replica  $E_b$ is:
\begin{eqnarray}
\label{speckle_backward}
{\mathcal E}_b(z,t,{\vec r}) \cong 
\exp[-i \omega t - i k_z z +i \theta_{_{Gs,Se}}(\tau)]
\cdot{f_{_{Gs,Se}}(\tau)}
&& \nonumber \\
\sum_{M}  a^{*}_{M} 
\cdot \exp[-i \vec{K}_M \vec r ].{\:}{\:}{\:}{\:}{\:}{\:}{\:}{\:}{\:}
\end{eqnarray}
The interference pattern inside PC mirror is given by 
\cite {Okulov:2010josa}: 
\begin{equation}
\label{twisted speckle}
I_{speckle}(z,t,{\vec r}) = |{\mathcal E}_f(z,t,{\vec r})
+{\mathcal E}_b(z,t,{\vec r})|^2.
\end{equation} 
The $3D$ distribution of intensity $I_{speckle}(z,t,{\vec r})$ 
is a random collection of pairs of 
helices with opposite handedness 
\cite {Okulov:2009} (fig.5). This feature of PC mirror ensures the 
phase-matched propagation of the phase-conjugated replica 
$E_b(z,t,{\vec r})$ \cite {Woerdemann:2009}.
The remarkable feature of this phase-conjugating laser 
interferometer technique \cite {Basov:1980, Okulov:2010josa} 
is a perfect compensation of the phase-piston 
errors $\Delta \phi_{mn}$ 
in backward propagation through the amplifying array 
\cite{Rockwell:1986}(fig.5). As a result 
the backwardly amplified beams will be collected in 
$beamsplitter$ $tree$ (fig.1b) in a single smooth beam identical to 
master oscillator output. The $\pi$-$shift$ between 
reconstructed $chirped$ backward waves which 
is necessary for decoupling from 
master oscillator $MO$ \cite {Basov:1980} may be 
produced by appropriate 
modulation inside $\lambda(t)/2$ Pockels cell (fig.1b).

The requirement of low PCM reflection $R_{_{PCM}}\cong 0.2$ imposes  
an additional link between backward gain in fiber array 
and energy 
of pump beams $E_1,E_2$ in PCM. According to IZEST-ICAN project 
\cite {Mourou:2013} the output energy between 
final $\bf BS$ and compressor (fig.1b) 
must be about $30-40 J$. For the realistic nanosecond 
spatially smooth beams $E_1,E_2$ 
with energy  $W_{PCM_{pump}} \sim 0.1 - 0.5 J$ the 
reflected chirped pulse energy 
cannot exceed $W_{PCM_{back}} \sim 0.02 - 0.1 J$. Consequently 
the backward gain in fiber network 
$G_{mn}=\sigma_{_{Yb}} \int_0^{L_{mn}} N_0(z^{'})d z^{'}$
should be large enough to reach the level of $30-40 J$. The 
reduction of gain $G_{mn}$ would require a more 
energetic pump beams $E_1,E_2$. 

\section{Temporal envelope deformation due to gain saturation.}

The nonlinear amplification of the 
light pulse in rare earth doped fiber \cite {Winful:2010,Davidson:2012}
modifies temporal envelopes. 
For incoherent amplification regime of short laser pulse
\cite {Basov:1966} 
$T_1>\tau_{_{Ga,Se}}>T_2$ in a presence of self-focusing 
the pulse propagation in each ${(m,n)}$  fiber with gain 
$G_{mn}=\sigma_{_{Yb}} \int_0^{{L_{mn}}} N_0(z^{'})d z^{'}$ is 
described by nonlinear Shrodinger-Frantz-Nodvik equation 
\cite {Okulov:1988_QE}: 
\begin{eqnarray}
\label{nanosec}
\ {\frac {\partial {{ E}_{f,b}}(z,t,\vec r )} {\partial z} } \pm 
{\frac {n_r} {c} }{\frac {\partial {{{ E}_{f,b}}}} {\partial t} }+
{\frac {i}{2 k_p}} {\nabla}_{\bot}^{{\:}2} {{ E}_{f,b}} =
&& \nonumber \\
{\frac { \sigma_{_{Yb}}{\:}{N{(z,t)}}} {2{\:}  } } 
{{ E}_{f,b}}+ i k n_2 |{{ E}_{f}+{ E}_{b}}|^2 \cdot {{ E}_{f,b}}, 
\end{eqnarray}
where $\sigma_{_{Yb}} \sim 10^{-20} cm^2$ is stimulated 
cross section of $Yb^{3+}$ resonant transition, $T_1, T_2$ are 
longitudinal and transversal relaxation times. The ultimate 
optical flux $F_{_{lim}}$ is limited by 
$F_{_{lim}} \sim  {\hbar \omega}/{2 \sigma_{_{Yb}} } $ 
\cite {Bulanov:2006}. The dynamics of population 
inversion $N(z,t)$ follows to rate equation:
\begin{equation}
\label{inversion nanosec}
\ {\frac {\partial N{(z,t)}} {\partial t}} =
- \sigma_{_{Yb}} N(z,t) \cdot |{{ E}_{f}+{ E}_{b}}|^2 +
{\frac { {N_0{(z)}-N{(z,t)}}} {T_1} },
\end{equation}
Indeed duration $\tau_{_{Gs,Se}} \cong 10^{-9}sec$ of CPA 
pulse \cite {Okulov:1988_QE} defines the incoherent 
dynamics of amplification $T_2 << \tau_{_{Gs,Se}} << T_1$ 
\cite {Basov:1966}. For a geometry of 
laser network under consideration (fig.1a,b) the average 
length of fiber amplifier set 
$<{L_{mn}}>=L_{f} \sim 1-10 {\:}meters$ exceeds  spatial length of pulse 
$c \cdot \tau_{_{Gs,Se}} \le 30 {\:}cm$ by an order of magnitude.  

Thus backward wave ${{ E}_{b}}$ is amplified without interference with 
forward wave ${{ E}_{f}}$:  
\begin{eqnarray}
\label{SS_amp}
{\frac {\partial {{ E}_{f,b}}(z,t )} {\partial z} } \pm 
{\frac {n_r} {c} }{\frac {\partial {{{ E}_{f,b}(z,t )}}} {\partial t} }
= {\frac { \sigma_{_{Yb}}{\:}{N_0{(z)}{{ E}_{f,b}(z,t )}}} {2}} \times
&& \nonumber \\
\exp [-2 \sigma_{_{Yb}} 
{\int_{_{_{-\infty}}}^{{\:}t}}|{{ E}_{f,b}}|^2 d \tau]
- i k n_2 |{{ E}_{f,b}(z,t )}|^2 {{ E}_{f,b}(z,t )},{\:}{\:}{\:}{\:} 
\end{eqnarray}

where integration is over the whole 
pulse prehistory $ - \infty< \tau < t$. 
As a result the envelope of backward pulse ${{ E}_{b}}$ 
is transformed 
due to gain saturation in accordance with 
exact solution \cite {Basov:1966,Okulov:1988_QE}: 
\begin{widetext}
\begin{equation}
\label{pulse}
{{ E}_{b}}(z,t )={\frac {{{ E}_{b}}
(L_{f},t )
\cdot exp{\:}[-i k n_2 \int_{_{_{-\infty}}}^{{\:}z} 
|{{ E}_{b}}(z,t )|^2 d z]}
{\sqrt{1-[1- exp{\:}[-\sigma_{_{Yb}} 
\int_{L_{f}}^{z} N_0(z^{'})d z^{'}]
{\:} exp{\:}[{\:}-2\sigma_{_{Yb}} 
\int_{-\infty}^{t-z/v_g}{{|{ E}_{b}}(z,\tau)|}^2 
d \tau]] } }}.
\end{equation}
\end{widetext}
The different regimes of pulse transformation are possible in this 
incoherent case. The most interesting feature is the 
self-similar regime 
of propagation which happens for initial conditions $E_{f,b}$ 
when both precursor and tail have an exponential form. 
This initial condition corresponds to 
specially formed hyperbolic $sech ((t- z n_0/c)/ \tau_{_{Se}})$ 
pulse of nanosecond duration. In this regime the maximum of 
intensity shifts toward the pulse 
precursor demonstrating seemingly 
superluminal propagation \cite {Basov:1966}. This regime is 
best suited for the chirp preservation. The exact temporal 
profile of phase-conjugated pulse for $sech^3(\tau)$ input pulse and 
$N_0(z)=N_0$ is given by:
\begin{widetext}
\begin{eqnarray}
\label{PC_sech_pulse}
{E_b}_{_{Se}}(z=0,\tau)=
{\frac {{{ E}_{b}}
(L_{f} )sech^3(\tau / \tau_{_{Se}})
\cdot exp{\:}[-i k n_2 \int_{_{_{-\infty}}}^{{\:}L_f} 
|{{ E}_{b}}(\tau )
sech^3(\tau / \tau_{_{Se}})|^2 d z]}
{\sqrt{1-[1- exp{\:}[-\sigma_{_{Yb}}N_0 
{L_{f}}  ]
{\:} exp{\:}[{\:}-2\sigma_{_{Yb}} {{ E}_{b}}^6(L_{f} )
[8+4sech^2(\tau / \tau_{_{Se}})+3sech^4(\tau / \tau_{_{Se}})]
tanh(\tau / \tau_{_{Se}})]] } }}
&& \nonumber \\.
\end{eqnarray}
\end{widetext}

Despite the speed 
of envelope maximum reaches $V \sim(6-10)\cdot c$ this regime 
of amplification 
does not violate causality. When the envelope maximum reaches 
a certain moment at precursor, which corresponds to 
oscillator threshold, 
the rectangular front moving with the speed $c$ appears.
\begin{figure}
\center{ \includegraphics[width=7  cm]{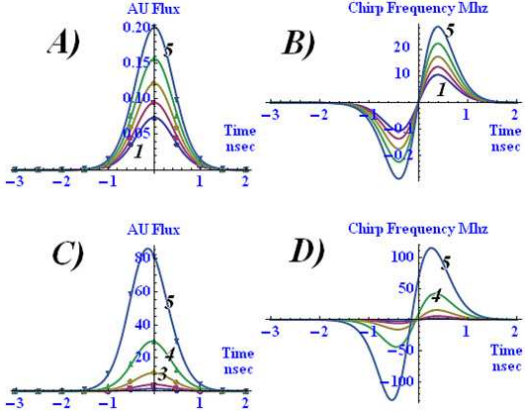}}
\caption{(Color online)Spectrum distortion of 
$2 \pi \delta \nu_{{_{Se}}}(\tau)= 
{\partial {\theta_{{_{Se}}}(\tau)}/{\partial \tau}}$ 
the chirped backward 
pulse 
$E_b \cong sech^3((t-z /v_g)/\tau_{_{Se}})$ with 
exponential precursor in 
backward amplifier before 
compressor . A),B) correspond to weak gain 
saturation ($G=2,2.5,3,3.5,4(1-5)$). C),D) stand for deep saturation 
regime ($G=6,7,8,9,10 (1-5)$).
The shift of the pulse maximum 
along precursor, i.e. to $negative$ $t$ corresponds 
to $superluminal$ propagation with envelope speed 
$V>c$ due to saturation of resonant transition.}
\label{fig.6}
\end{figure}
\begin{figure}
\includegraphics[width=7 cm]{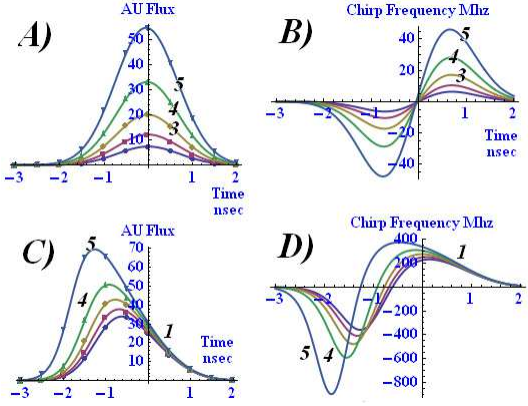}
\caption{(Color online)Spectrum distortion 
$2 \pi \delta \nu_{{_{Gs}}}(\tau)= 
{\partial {\theta_{{_{Gs}}}(\tau)}/{\partial \tau}}$ 
of the chirped backward Gaussian pulse 
$E_b \cong exp{\:}[-3(t-z /v_g)^2/2 \tau_{_{Ga}}]$ in 
backward amplifier before 
compressor.  A),B) correspond to weak gain 
saturation ($G=2,2.5,3,3.5,4 (1-5)$). C),D) stand for deep saturation 
regime ($G=6,6.1,6.2,6.3,6.4(1-5)$).The shift of the Gaussian 
pulse maximum 
along precursor and temporal steepening corresponds  
to $subluminal$ propagation with envelope  speed 
$V<c$.}
\label{fig.7}
\end{figure}
Fig.(6 a,b) 
shows the deformation of $E_b(z=0, \tau) \sim sech^3(\tau)$ 
temporal profile for small signal 
gain increments in the range $G=\sigma_{_{Yb}}N_0 {L_{f}}\cong 2-9$. 
The deep saturation shifts the pulse maximum and chirp 
towards precursor for a several hundreds of picoseconds in 
a self-similar way (Fig.6c,d). The asymmetry induced by gain 
saturation is weak due exponential precursor. The chirp 
varies smoothly for tens of megahertz. 

The other initial condition for $E_b(z=0, \tau)$ 
corresponds to Gaussian $exp(-\tau^2 / \tau^2_{_{Ga}})$  
nanosecond pulse. The exact temporal 
profile of phase-conjugated pulse for 
$exp^3(-\tau^2 / \tau^2_{_{Ga}})$ input pulse is:
\begin{widetext}
\begin{equation}
\label{PC_gauss_pulse}
{E_b}_{_{Ga}}(z=0,\tau)={\frac {{{ E}_{b}}
(L_{f} )\exp(-3\tau^2 / \tau^2_{_{Ga}})
\cdot exp{\:}[-i k n_2 \int_{_{_{-\infty}}}^{{\:}L_f} 
 |{{ E}_{b}}(\tau )
exp(-3 \tau^2 / \tau^2_{_{Ga}})|^2 dz ]}
{\sqrt{1-[1- exp{\:}[-\sigma_{_{Yb}}N_0 
{L_{f}}  ]
{\:} exp{\:}[{\:}-\sigma_{_{Yb}} \sqrt{\pi}{{ E}_{b}}^6(L_{f} )
[erf (6 \tau / \tau_{_{Ga}})]]] } }},
\end{equation}
\end{widetext}
where $erf(6 \tau / \tau_{_{Ga}})$ is 
$\int_{-\infty}^{\tau } 
\exp[- 6 {\tau^{'}}^2/ \tau^2_{_{Ga}}] d \tau^{'}$.
The deformation of 
${E_b}_{_{Ga}}(z=0, \tau) \sim \exp^3(-\tau^2 / \tau^2_{_{Ga}}))$ 
temporal profile for small signal 
gain increments in the range $G=\sigma_{_{Yb}}N_0 {L_{f}}\cong 2-6$ 
is shown at Fig.7. 
The deep saturation also shifts the pulse maximum and chirp 
along precursor. But temporal shift is more than a nanosecond 
and asymmetry induced by gain 
saturation is much stronger. The pulse is shortened and chirp 
varies in a range of the hundreds of megahertz. 

\section{Discussion. Figure of merit for a phase-locked output.}

For the time modulated carrier frequency $\omega$ the interference 
pattern of the each pair 
of the fiber 
amplified waves 
inside phase-locking Mach-Zehnder and Michelson interferometers 
is characterized by 
effectiveness of  beam combination taking into account 
random phase-piston errors 
\cite {Tunnermann:2011,Galvanauskas:2012}. Suppose that beams are 
perfectly overlapped at beamsplitter and phase piston 
error is almost compensated. 
Let us assume 
that beams $E_{1,2}(\omega)$ have identical spectral 
power $P_0(\omega)$ \cite {Rytov:1987}. The remained 
phase-jitter $\Delta \Phi_{1,2}(\omega)$ 
is due to chirp deformation in amplification. The effectiveness 
is measured by figure of merit (FOM) which 
is defined as a visibility of interference pattern at beamsplitter 
$\bf BS$ 
for each given spectral 
harmonic \cite {Tunnermann:2010, Brignon:2011}:

\begin{eqnarray}
\label{fom}
FOM_{12} (\omega )=
\frac {P_{comb}(\omega)-P_{idle}(\omega)}
{P_{comb}(\omega)+P_{idle}(\omega)} 
&& \nonumber \\
= {\frac {|E_1 (\omega)+E_2 (\omega) 
\cdot \exp{(i \Delta \Phi_{1,2}(\omega))}|^2}{|E_1 (\omega)|^2 + 
|E_2 (\omega)|^2}},
\end{eqnarray}
where $P_{comb}(\omega)$ is a spectral component of optical flux 
measured in output port of beam splitter, 
$P_{idle}(\omega)$ is a spectral power in idle port. 
For the above assumption of equal spectral density:
\begin{eqnarray}
\label{fom_two_beams}
FOM_{12} (\omega )
= {\frac {|E_1 (\omega)+E_2 (\omega) 
\cdot \exp{(i \Delta \Phi_{1,2}(\omega))}|^2}{|E_1 (\omega)|^2 + 
|E_2 (\omega)|^2}}=
&& \nonumber \\
\pm \frac {2 P_0(\omega) \cos[\Delta \Phi_{1,2}
(\omega)]}{2 P_0(\omega)}=\pm  \cos [\Delta \Phi_{1,2}(\omega)].
{\:}{\:}{\:}{\:}{\:}{\:}{\:}{\:}
\end{eqnarray}

The figure of merit for all spectral components is evaluated 
as integral over all frequencies: 
\begin{eqnarray}
\label{fom_two_beams_total}
FOM_{12}= C \cdot \int s(\omega)
FOM_{12}(\omega) d \omega = 
&& \nonumber \\
\pm {\:}C \cdot \int s(\omega)
\cos [\Delta \Phi_{1,2}(\omega)] d \omega , {\:}{\:}
C = 1 / \int s(\omega)d(\omega),{\:}{\:}{\:}{\:}{\:}{\:}
\end{eqnarray}
where $s (\omega)$ is normalised spectral intensity.

The FOM for the $N_f=2^{N_{ex}}$ beams coherently 
added to each other at all components of 
the  $binary{\:}{\:} BST$ 
(fig.1) may be evaluated as interference pattern of 
$2^{N_{ex}}$ beams with fluctuating 
phases $\Delta \Phi_{n}(\omega)$. When phase piston errors 
$\Delta \phi_{mn}$ are 
eliminated by precise adjustment in Mach-Zehnder scheme or 
phase-conjugation in Michelson scheme, the result of 
constructive interference at output beamsplitter $\bf B$ 
reads as: 
\begin{eqnarray}
\label{output BS}
E_b(z=BS,\omega) = {\frac{1}{\sqrt{N_f}}}
\sum_{n}^{N_f}  E_{0}(\omega) \exp [i \Phi_n (\omega)]=
&& \nonumber \\
E_{0}(\omega_0){\sqrt{\frac{s(\omega)}{{N_f}}}}
\sum_{n}^{N_f}   \exp [i \Phi_n (\omega)].{\:}{\:}{\:}{\:}{\:}{\:}
\end{eqnarray}
This gives the spectral intensity $I(\omega)$ at output as: 
\begin{eqnarray}
\label{spectral intensity output BS}
I(\omega)=E_b(\omega) E_b^{*}(\omega)= 
 |E_{0}(\omega)|^2 {\frac{s(\omega)}{{N_f}}}[ N_f +
&& \nonumber \\
\sum_{n,m \ne n}^{N_f} \sum_{n}^{N_f}  
\cos [i \Phi_{mn} (\omega)]],
\end{eqnarray}

 where $\Phi_{mn} (\omega)$ is relative phase fluctuation between 
 $m$ and $n$ fiber channel at frequency $\omega$. The figure of merit 
 over all $\omega$ for $N_f$ channels is given by:
\begin{eqnarray}
\label{FOM_array}
FOM={\frac {\int (I(\omega)-(N_f\cdot s(\omega) - 
I(\omega))) d \omega}{\int N_f \cdot s(\omega)d \omega}}= 
{\frac{C}{{N_f}}} \times 
&& \nonumber \\
 \int  s(\omega)[ (2-N_f) +
{\frac {2}{N_f}}\sum_{n,m \ne n}^{N_f} 
\sum_{n}^{N_f}  
\cos [i \Phi_{mn} (\omega)]] d \omega,{\:}{\:}{\:}{\:}{\:}{\:}
\end{eqnarray}
where substitution of relative $FOM_{mn}$ between two 
channels gives:
\begin{equation}
\label{FOM_algebraic}
FOM=[{\frac {2}{N_f}}-1]+{\frac {2}{N^2_{f}}} 
\sum_{n,m \ne n}^{N_f} \sum_{n}^{N_f} \cdot FOM_{nm}.
\end{equation}
Final evaluation may be simplified else under assumption 
$FOM_{nm}=FOM_{12}$ \cite {Tunnermann:2011}:
\begin{equation}
\label{FOM_homogeneous_array}
FOM_{large} \cong [{\frac {2}{N_f}}-1]+
2[1-{\frac {1}{N_{f}}}]\cdot FOM_{12}.
\end{equation}
In the framework of above formulated model the 
asymptotic behavior ($large {\:} N_f $) of array 
network with small interchannel phase fluctuations 
$\Phi_{mn}(\omega)$ demonstrates the tendency to 
FOM of a single pair of channels \cite {Tunnermann:2011}.
 
\section{Conclusions}
In this work the theory of 
phase-locking configurations of the fiber amplifying network 
composed of $N_f$ lasers is presented. It is shown that a single pass 
$binary{\:}{\:} BST$ tree Mach-Zehnder combiner is sensitive to variations of 
$N_{var}=(N_f-1)*6 + 2*(N_f-1)*(2+2+1)$ degrees 
of freedom. The growth of $N_{var}$ is linear with respect to 
size of fiber array. On the other hand the double 
pass Michelson configuration 
with DFWM phase-conjugating mirror automatically adjusts backwardly 
reflected signal $E_b$ with fibers and 
$N_{BS}$ beamsplitters. The essential condition of proper 
PCM operation 
is that information about fiber network stored within PC 
mirror dynamical hologram \cite{Soskin:1990} 
should be large enough to remember layout of the 
entrance  $binary{\:}{\:} BST$ and distribution of the 
phase-pistons $\Delta \phi_{mn}$. This issue is tightly 
connected with the phase conjugation fidelity ( pump / signal 
correlation  $K_{_{PCM}}$)  whose high value 
($\ge 0.9$) \cite {Basov:1980} is experimentally compatible 
with low PC reflectivity $R_{_{PCM}} \sim 0.2$. 

Preliminary evaluation shows that proper energy efficiency 
of the Michelson 
configuration (fig.1b) 
requires weakly saturated gain of forward pulse ${E_f}(\tau)$ up 
to $\sim 100 \mu J$ with almost immediate (within $\sim 10 ns$) 
amplification of the 
backward pulse ${E_b}(\tau)$ from $\sim 20 \mu J$ up to 
$\sim 2-3 mJ$. These $\sim 0.8 J$ 
per pulse losses are small compared to heating losses 
in fiber CFA evaluated at the level of 10-20 percents 
\cite {Mourou:2013}. As is shown already \cite {Eidam:2011} 
the saturated gain for backward pulse 
could provide $2.2 mJ$ level output per each 
large mode area photonic crystal fiber amplifier. 
Thus smooth spatial mode $30-40 J$ chirped pulse output before 
compressor looks quite realistic.
 
In Michelson interferometric configuration the phase-piston 
errors of the each pair of beams 
$\Delta \phi_{mn}$ are compensated via phase-conjugating 
action of DFWM mirror \cite {Basov:1980}. On the contrary 
the Mach-Zehnder scheme 
is highly sensitive to variations of optical path's and orientations 
of beamsplitters \cite {Goodno:2010}. The optimization of the figure of 
merit requires 
the reasonable balance of gains $G_{mn}$  inside 
fiber amplifier's network in order to avoid a randomization  
of the maxima of temporal envelopes and corresponding random 
displacement of frequency chirp $\delta \omega_{mn}(t)$ in 
a set of fiber amplifiers. The gain values in Michelson 
fiber network 
$G_{mn}$ are tightly linked with reflectivity of 
PC mirror and energy of DFWM PCM pump beams. 

In both cases the distortions of temporal profile 
${E_b}_{_{Ga,Se}}(z=0,\tau)$ seriously 
affect the frequency chirp. The $sech$ envelope is less sensitive 
to nonlinear distortion in fiber amplifier with saturated gain 
because of exponential precursor. There is a definite 
range of fiber gain $G_{mn}$ where the pulse maximum moves 
towards the precursor with superluminal speed with sufficiently 
small displacement of the chirp towards precursor. 
The Gaussian temporal profile demonstrates much stronger deformation 
of the envelope due to saturation of the fiber 
gain $G_{mn}$ because Gaussian precursor decelerates pulse 
maximum and the envelope ${E_b}_{_{Ga}}(z=0,\tau)$ 
becomes steeper.  

\section{Acknowledgments}
The profound gratitude is expressed to Prof.G.Mourou 
for the stimulating discussion on 
compatibility of chirped pulse amplification 
technique with stimulated Brillouin scattering 
optical phase-conjugation and support 
to attend an IZEST-ICAN workshop in CERN on June 27-28, 2013. 
The author wishes to acknowledge 
Prof. A.Brignon and Prof. J.Limpert for discussions 
of phase synchronization of the 
tightly packed fiber optical amplifiers.

\end{document}